\documentstyle[epsf,12pt]{article}
\def\2pi4{(2\pi)^4}

\begin{document}

\centerline{\Large \bf The nuclear binding and the EMC effect in the deuteron}

\vspace*{4mm}

\centerline{L.P. Kaptari}

\vspace*{-4.5mm}

\begin{center}
{\small \em Bogoliubov  Laboratory of Theoretical Physics,\\
 Joint Institute for Nuclear Research, 141980, Dubna, Russia}
\end{center}\vspace*{-4mm}

\centerline{K.Yu. Kazakov
\footnote{On leave from Laboratory of Theoretical 
and Nuclear Physics, Far Eastern State University, Vladivostok,
690000, Russia}}

\vspace*{-4.5mm}

\begin{center}
{\small \em The Niels Bohr Institute, Blegdamsvej 17, DK-2100 Copenhagen \O,
Denmark}
\end{center}

\vskip 2mm 

\begin{abstract}
Influence of the nuclear binding in the deuteron deep inelastic structure 
function $F_2(x,Q^2)$ is investigated. The description is based on
the Bethe-Salpeter formalism in the ladder approximation 
and the operator product expansion within an
effective meson-nucleon theory. It is shown that the binding in a
fully relativistic treatment can be simply parametrised and its origin
can be understood in terms of conventional nuclear degrees of
freedom. A numerical estimate of the EMC effect in the deuteron is given. 
\end{abstract}

\section{Introduction}
\indent

Ever since nuclear effects in deep inelastic lepton scattering on hadron
targets have clearly manifested itself over the last decade, both particle
and nuclear theorists obtained a profound insight into physics of the EMC
effect. However  a majority of models of the EMC effect provided by a fully
relativistic hadronic theory resorts to the Relativistic impulse
approximation (RIA) with the off-mass-shell kinematics, for example see
Refs.~\cite{gross,umnikov}.

It seems to be evident that the free nucleon structure function (SF)  
$F_2^N(x,Q^2)$ is somehow affected by the nuclear medium. Here $Q^2$ and 
$x$ are the four momentum transfer squared and the Bjorken scaling variable 
$x=Q^2/(2m\nu)$, with $\nu$ being the energy transfer in the
laboratory frame. Conventional nuclear physics models, for 
a review see Ref.~\cite{arneodo}, assume that properties of nuclear 
constituents, nucleons and mesons, are insensitive to the nuclear
environment. Yet the hadronic  
current, which is poorly understood object so far, may depend on 
dynamics of nucleons inside a nucleus and their interactions. The common
assumption in this case is to treat the hadronic current as the sum of
currents of single (quasi)free nucleons, whereas the relation
between the momentum and energy of the virtual nucleon is specified in
a relevant treatment of the relativistic bound state. Such a proviso 
substantiates derivations of the convolution formula for the nuclear SF
with a consideration of relativistic kinematics. 

The approach we take in this paper follows a rigorous method 
introduced in Refs.~\cite{new,physrev} to investigate deep inelastic
scattering on light nuclei. The method relays on the leading order 
operator product expansion (OPE) within the effective meson-nucleon
theory. It allows one to derive the convolution formula for the
nuclear SF which 
accounts for the role of the Fermi motion and binding effects in a 
new and unambiguous form. Such a formulation yields the $n$th moment
of the nuclear SF as a product of the moments of the isolated-hadron
SF's and the moments of the effective hadron momentum 
distributions. The latter
are given by matrix elements of twist-two operators sandwiched between
 the ground state vector of the target. 

Using the Hamiltonian formalism relevant operators in the
OPE are found in the lowest order approximation on the
meson-nucleon coupling constant and resulting operators 
involve the meson degrees of freedom. Finally the 
matrix elements of these operators 
  are calculated over the ground state of the target in the non-relativistic 
(NR) limit. The
coefficient functions together with meson cloud contributions are 
substituted by the moments of the isolated-hadron SF.     

In this paper we apply the method proposed in
Refs.~\cite{new,physrev} in order to describe the deuteron SF $F_2^D(x,Q^2)$ 
where the system is viewed as the relativistic 
bound state in the Bethe-Salpeter (BS) formalism. To this end
expressions for the twist-two operators in the OPE 
are found by making use of the
relativistic Hamiltonian. Explicit expressions for the moments of 
the deuteron SF are found through actual calculation of the matrix 
elements in terms of the BS vertex function. At the first step we
assume that the off-shell effect on quark and 
anti-quark momentum distributions in hadrons may be 
neglected\footnote{Though it may well be incorporated in this line
of considerations.}. In this way the Mellin transform reconstructs
the deuteron SF as a sum of convolution terms describing the Fermi
motion and the binding in the deuteron due to meson exchanges.
Since meson exchange currents (MEC) are associated with exchange 
mechanisms responsible for the binding in the deuteron, momentum
distribution for the constituent mesons reveal itself as well.  

Advantage of this relativistic approach is that it  
differentiates the pure binding effects from the off-shell effects as
both have an unlike origin. The magnitude of the former is given in 
terms of well-understood quantities such as the kinetic energy of a
nucleon and the binding energy of the deuteron. Besides dealing with
the deuteron we can obtain all required formulae analytically within 
well-defined approximations and subject the problem to intense scrutiny.  

The paper is divided into five sections. In Sec. II the basic 
formalism is reviewed. The twist-two operators in the OPE of
the Compton amplitude is specified. In Sec. III the moments of the
fractional momentum distributions of nucleons and mesons in the deuteron are 
calculated. The NR limit of the energy-momentum sum rule is performed. 
Computation  of $F_2^D(x,Q^2)$ as a function of $x$ is carried out in 
Sec. IV.  And Sec. V contains a summary of principle results and 
conclusions.

\section{Structure functions of the deuteron in the BS approach}
\subsection{The relativistic vertex function of the deuteron}
\indent

Consider a system of two nucleons interacting via a scalar, pseudo-scalar
and vector meson exchanges. The Lagrangian density for meson-nucleon 
couplings is
\begin{eqnarray}
{\cal L}_I=-g_V\bar N\gamma_\mu N V^\mu  - g_S\bar N  N S  -\imath g_\pi
\bar N\gamma_5\tau_i N\varphi_i +\ldots \label{lag}
\end{eqnarray}
with $N$ being the nucleon and $S$, $\varphi_i$ and $V_\mu$  meson 
fields.

Assuming that the NN interaction to 
proceed through the meson
exchange only, we start with the one-boson-exchange (OBE) model  
in application to the BS
equation (BSE) in the bound state region of the deuteron. Therefore the
BS amplitude, $\chi(p;P)=\langle 0|
T N_{p_1} N_{p_2}|P\rangle$, and equivalently the BS  vertex function 
$\Gamma(p;P)=S^{-1}(p_1,p_2)\chi(p;P)$, satisfy the BSE in the
momentum space as 
\begin{eqnarray}
&&\Gamma(p;P)=\imath 
\int\frac{d^4p^\prime}{(2\pi)^4}{\cal V}_{\rm OBEP}(p-p^\prime)
\chi(p^\prime;P),\label{BS-eq}
\end{eqnarray}
where $S(p_1,p_2)$ is the free two-nucleon Green's function and 
${\cal V}_{\rm OBEP}$ is the OBE interaction kernel. 
The BSE~(\ref{BS-eq}) is
written  in terms of the relative momenta $p$, $p^\prime$ of the 
nucleons and the center-of-mass (c.m.) momentum $P$ 
and $p_{1,2}=P/2\pm p$. The interaction kernel ${\cal V}_{\rm OBEP}$ is 
the sum of one-particle exchange amplitudes of certain bosons with given
masses and couplings generated by ${\cal L}_I$ in Eq.~(\ref{lag}).  

Since the BSE is homogeneous, it cannot determine a multiplicative
constant of $\chi(p;P)$. In order to normalise the BS amplitude,
condition based on the electro-magnetic (EM) current conservation at zero 
momentum transfer can be used. In the IA the normalisation condition reads 
\begin{eqnarray}
-\frac{\imath}{N}\int d^4p\, 
\bar\chi(p;P)
[\frac{\partial}{\partial P_0}S^{-1}(p_1,p_2)]\chi(p;P)=P_0,
\label{normcon}
\end{eqnarray}
where $P_0=\sqrt{{\bf P}^2+M^2}$ and 
$\bar\chi$ stands for the BS
amplitude  conjugate of $\chi$. An appropriate normalisation constant
in Eq.~(\ref{normcon}) is denoted as $N$. In the Euclidean space the 
amplitude $\chi$  and its conjugate $\bar\chi$ are related as follows:
$\bar\chi(p_4,{\bf p};P)=-[\chi(-p_4,{\bf p};P)]^*\gamma_0^{(1)} 
\gamma_0^{(2)}$, where $p_4$ lies on the imaginary axis.

The Lagrangian~(\ref{lag}) and the BSE~(\ref{BS-eq}) define an effective
meson-nucleon theory for the description of the bound state of the 
deuteron which is meaningful in a low-order approximation.
Such a theory is not fundamental one in a sense that interactions in 
Eq.~(\ref{lag}) are viewed as effective ones which are 
modified by vertex form factors. If the BSE is applied in the ladder 
approximation as done in Eq.~(\ref{BS-eq}), all 
further calculation of matrix elements of EM currents  
expressed in terms of the nucleon and meson fields over the deuteron
state have to be made up to the second order in the strong coupling. 
This procedure is justified  in describing the nuclear structure
effects in electron scattering.

\subsection{Operator product expansion}
\indent 

The structure functions $F_{1,2}$ can be
calculated via the OPE of the amplitude 
for forward Compton scattering of virtual photons off a hadronic
target~\cite{wilson,brandt}. The expansion separates characteristic
hadronic and nuclear scales and formulates this 
particular analysis of the nuclear SF
as a calculation of matrix elements of certain operators within low-energy 
physics.   
 
The spin-averaged Compton amplitude is expanded in terms of local 
operators $O$ and coefficient functions $E$ in the form 
\begin{eqnarray}
T_{\mu\nu}(P,q)\propto& -&g_{\mu\nu}\sum\limits_{n=even}
\left(\frac{2}{Q^2}\right)^nq_{\mu_1}\ldots q_{\mu_n}
\sum\limits_{a=N,B} E^n_{1,a}(Q^2,g)
\langle P|O^{\mu_1\ldots\mu_n}_{a}|P\rangle\nonumber \\
&+&2g_{\mu\mu_1}g_{\nu\mu_2}\sum\limits_{n=even}
\left(\frac{2}{Q^2}\right)^{n-1}q_{\mu_3}\ldots q_{\mu_n}
\sum\limits_{a=N,B} E^n_{2,a}(Q^2,g)
\langle P|O^{\mu_1\ldots\mu_n}_{a}|P\rangle,\label{compton}
\end{eqnarray}
where $O^{\mu_1\ldots\mu_n}_a$ are series of operators having twist two 
at the level of naive dimension counting and 
$a$ tags the fundamental fields of the theory, i.e. nucleons (N) and 
bosons (B). Each operator is
appropriately symmetrised and trace-subtracted. In this procedure the 
deuteron SF $F_2$ is expressed in terms of invariant
diagonal matrix elements $\mu^{a/D}_n$
\begin{equation}
\langle P|O^{\mu_1\ldots\mu_n}_a|P\rangle=
\{P^{\mu_1}\ldots P^{\mu_n}\}
\mu^{a/D}_n,
\end{equation} 
and the coefficient functions $E^n_{2,a}$ by 
\begin{equation}
{\sf M}_{n-1}(F_2^D)=\sum_{a=N,B} E^{(2)}_{a,n}(Q^2,g)\mu^{a/D}_n,
\label{mom-conv}
\end{equation}
where ${\sf M}_n(F)$ defines the $n$th moment of $F$, 
${\sf M}_n(F)=\int_0^1dxx^{n-1}F(x,Q^2)$. In Eq.~(\ref{mom-conv})
we neglect the effects of nucleon and target masses corrections  
which can be taken into account in a closed analytic form. The
explicit line of our calculation proceeds as follows: (1) Define a basis
of the twist-two operators in OPE~(\ref{compton}). (2) Parametrise
the coefficient functions in terms of moments
of SF's $F_2$ of nucleons and bosons, $E^n_{2,a}\propto {\sf
M}_{n-1}(F_2^a)$. According to our assumption  $E^n_a$ are 
identified with the corresponding moments of a  
measured  isolated-hadron SF, which in case of $a=N$  include
effects of meson clouds surrounding a nucleon.  (3) Obtain the moments
of the deuteron SF by
calculating the matrix elements  $\mu^{a/D}_n$ in 
Eq.~(\ref{mom-conv}). (4) Use the Mellin transform to invert the
moments and exhibit SF's. 

Let us determine explicitly the operators in the expansion
(\ref{compton}) for the meson-nucleon theory. For a nucleon field and
a pseudo-scalar (and  scalar)
field interacting via coupling given by (\ref{lag}) there are only two series
of symmetric bilinear operators 
\begin{eqnarray}
&&{\cal O}_{N}^{\mu_1\ldots\mu_n}
=\left(\frac{\imath}{2}\right)^{n-1}
{\cal S}\{:\bar N\gamma^{\mu_1}
\stackrel{\leftrightarrow}{\partial}^{\mu_2}\ldots
\stackrel{\leftrightarrow}{\partial}^{\mu_n}N:\},\label{sca-N}\\
&&{\cal O}_{\pi}^{\mu_1\ldots\mu_n}
=\left(\frac{\imath}{2}\right)^{n-1}
{\cal S}\{:\varphi_i
\stackrel{\leftrightarrow}{\partial}^{\mu_1}\ldots
\stackrel{\leftrightarrow}{\partial}^{\mu_n}\varphi_i:\}\label{sca-S},
\end{eqnarray}
where ${\cal S}$ implies complete symmetrization in the Lorentz indices
$\mu_1\ldots\mu_n$ and removes all traces with respect to each pair
of indices.  

If a theory involves fundamental massive vector bosons, matters become more 
complicated by the fact that the vector field itself has twist zero
and some subsidiary condition is needed to introduce the set of 
relevant operators.  Two sets of the operators can be formed.

Set (a): 
\begin{eqnarray}
&&{\cal O}_{N}^{\mu_1\ldots\mu_n}
=\left(\frac{\imath}{2}\right)^{n-1}
{\cal S}\{:\bar N\gamma^{\mu_1}
\stackrel{\leftrightarrow}{D}^{\mu_2}\ldots
\stackrel{\leftrightarrow}{D}^{\mu_n}N:\},\label{vec-N}\\
&&{\cal O}_{F}^{\mu_1\ldots\mu_n}
=\left(\frac{\imath}{2}\right)^{n-2}
{\cal S}\{:F^{\lambda\mu_1}
\stackrel{\leftrightarrow}{\partial}^{\mu_2}\ldots
\stackrel{\leftrightarrow}{\partial}^{\mu_{n-1}}F_\lambda^{
\phantom{\nu}\mu_n}:\}
\label{vec-V1},\\
&&{\cal O}_{V}^{\mu_1\ldots\mu_n}
=\left(\frac{\imath}{2}\right)^{n-2}\mu_V^2
{\cal S}\{:V^{\mu_1}
\stackrel{\leftrightarrow}{\partial}^{\mu_2}\ldots
\stackrel{\leftrightarrow}{\partial}^{\mu_{n-1}}V^{\mu_n}:\},
\label{vec-V2}
\end{eqnarray} 

and set (b): 
\begin{eqnarray}
&&{\cal O}_{N}^{\mu_1\ldots\mu_n}
=\left(\frac{\imath}{2}\right)^{n-1}
{\cal S}\{:\bar N\gamma^{\mu_1}
\stackrel{\leftrightarrow}{\partial}^{\mu_2}\ldots
\stackrel{\leftrightarrow}{\partial}^{\mu_n}N:\},\label{vec-N1}\\
&&{\cal O}_{V}^{\mu_1\ldots\mu_n}
=\left(\frac{\imath}{2}\right)^{n-1}
{\cal S}\{:V_\lambda
\stackrel{\leftrightarrow}{\partial}^{\mu_1}\ldots
\stackrel{\leftrightarrow}{\partial}^{\mu_n}V^\lambda:\}\label{vec-V3}.
\end{eqnarray}

Set (a) is introduced on the ground 
of gauge-invariance~\cite{christ}. The first two 
operators labelled by $N$ and $F$ are invariant under the 
transformation $V_\mu(x)\to V_\mu(x)+g_V\partial_\mu\Lambda(x)$ and 
$N(x)\to \exp{(\imath g_V\Lambda(x))}N(x)$, although the mass term in 
the free Lagrangian for the vector field $-\frac12 \mu_V^2V^2$ breaks
this gauge symmetry; and the subsidiary 
condition is used~\cite{birbrair} that
the symmetrised form of the energy-momentum tensor for the massive
vector bosons is restored by the operators when $n=2$: 
\begin{eqnarray}
\Theta^{\mu\nu}=F^{\lambda\mu}F_{\lambda}^{\phantom{\mu}\nu}
+\mu_V^2V^\mu V^\nu - 
\frac12 g_V(\bar N\gamma^\mu N V^\nu+\bar N\gamma^\nu N
V^\mu)-g^{\mu\nu}
{\cal L}.
\end{eqnarray}
The operators $O_V^{\mu_1\ldots\mu_n}$~(\ref{vec-V3}) guarantee that. 
On the other hand set (b) is chosen on the basis of that the  
Lagrangian of the vector field is taken in the form of a covariant sum
of four 
Lagrangians each separately corresponding to the components of $A_\mu$ and 
imposing the invariant subsidiary condition $\partial\cdot A=0$.
Thus the operators depending explicitly on the 
interaction and composed out of the tensor $F_{\mu\nu}$ are excluded
from the set. In this case the energy-momentum tensor is restored in the form 
\begin{eqnarray}
\Theta^{\mu\nu}=\partial^\mu A_\lambda \partial^\nu A^\lambda
-g^{\mu\nu}{\cal L}^\prime\label{tensor}
\end{eqnarray}
and the operators (\ref{vec-N1}) and (\ref{vec-V3}) preserve  
the momentum sum rule for the $F_2(x)$, i.e. ${\sf M}_1(F_2)=1$.
 
\section{The moments of the deuteron structure function}
\subsection{Nucleon contribution}
\indent 

The set (a) was considered in Ref.~\cite{birbrair} in a calculation of
the nuclear SF
within the Walecka model. The result is that the exchange by vector 
meson is incompatible with the observed magnitude of the EMC
effect. The nucleon part of the nuclear SF, because of the final-state 
interaction with the vector meson field of a nucleus, is found to be
twice as large as the experimentally observed one. Moreover the meson 
part peaks at small values of $x$, which is either not observed. 

We consider the twist-two operators that belong to the set (b). In
this set operators for the vector and scalar meson fields are taken
into account on equal basis. Taking the
Bjorken limit, defined by $Q^2,\nu\to\infty$ with $x$
fixed, and choosing a coordinate system so that $q=(\nu,0_\perp,q_l)$,
where $q_l=-\sqrt{Q^2+\nu^2}$, leads to the following. The
mathematical part is to compute the
$n$th $\partial_-$ derivatives, $q\cdot \partial=\nu\partial_-$,
 in Eqs.~(\ref{sca-N}), (\ref{sca-S}) and 
(\ref{vec-N1}) and (\ref{vec-V3}) by making use of the equation of
motion for the coupled meson and nucleon fields. Consequently the resulting 
operators involve the meson degrees of freedom. The matrix elements of
the output operators over the
deuteron state can be calculated by the Mandelstam
method~\cite{mand}.  

A representation
of the moments ${\sf M}_n(F_2^D)$ in terms of diagrams 
is shown on the Fig.~1, where the
inner solid lines denote the nucleon propagators and the wavy lines denote 
the meson propagators. The closed triangle and square indicates the 
corresponding moments of the isolated-hadron SF. The moments of
effective nucleon 
distributions, $\mu_n^{N/D}$, are described by Fig.~1(a) and (b) and
can be written down as 
\begin{eqnarray}
\mu_n^{N/D}=\mu_n^{N/D}({\rm RIA})+\mu^{N/D}_n({\rm int}),\label{mund}
\end{eqnarray}
where  we call the first term in Eq.~(\ref{mund}) as the RIA  with the 
on-mass-shell kinematics\footnote{Below  the normalisation constant
$N$ is included in the definition of the BS amplitude.}
\begin{eqnarray}
\mu_n^{N/D}({\rm RIA})=-\frac{\imath}{P_+}\int d^4p\,\bar\chi(p;P)
\gamma^{(1)}_+ S^{-1}(p_2)\chi(p;P)
(\frac{\tilde p_+}{P_+})^{n-1},\label{mu-IA}
\end{eqnarray}
where $S^{-1}(p_2)=m-\hat p_2$ and the $+$-component of a vector 
defined as the sum of the time and longitudinal components originates 
from $p\cdot q\approx \nu p_+$; $\tilde p$ is on-mass-shell
momentum, $\tilde p^2=m^2$.  The form of Eq.~(\ref{mu-IA}) is close to that
of Ref.~\cite{alexm}. Although the nucleon of the upper-half circle at
Figs.~1(a) and~1(b) is  confined kinematically 
to the mass-shell, its propagation 
is described in a covariant way as given in Eq.~(\ref{mu-IA}).

The contribution $\mu^{N/D}_n({\rm int})$ in Eq.~(\ref{mund}) is due to the 
strong interaction between two nucleons in the deuteron and is described by 
Fig.~1(b). It is obtained
through the calculation procedure of the operators~(\ref{vec-N}) in
the Hamiltonian approach:
\begin{eqnarray}
&&
\mu^{N/D}_n({\rm int})=\label{mu-int}\\
&&-
\frac{1}{P_+}\int \frac{d^4pd^4p^\prime}{(2\pi)^4}
\,\bar\chi(p;P)
\bigl[\gamma_+\gamma^0\bigr]^{(1)}{\cal V}_{{\rm
OBEP}}(k)\chi(p^\prime;P)
\frac{
\tilde p_+{}^{n-1}-
(\frac{\tilde p_+ +\tilde p^\prime_+ - k_+}{2})^{n-1}}{P_+^{n-1}
(\tilde p_+ -\tilde p^\prime_+ + k_+)}
\nonumber \\
&&-
\frac{1}{P_+}\int \frac{d^4pd^4p^\prime}{(2\pi)^4}\,
\bar\chi(p;P)
{\cal V}_{{\rm OBEP}}(k)\bigl[\gamma^0\gamma_+\bigr]^{(1)}
\chi(p^\prime;P)
\frac{\tilde p^\prime_+{}^{n-1}
-(\frac{\tilde p^\prime_+ +\tilde p_+ + k_+}{2})^{n-1}}
{P_+^{n-1}(\tilde p^\prime_+- \tilde p_+-k_+)}\nonumber,
\end{eqnarray}
where $k=p^\prime-p$ and $\tilde p$, $\tilde p^\prime$ are the
on-mass-shell nucleon momenta. 

\begin{figure}[t]
\input epsf
\epsfxsize=14cm
\centerline{\epsfbox{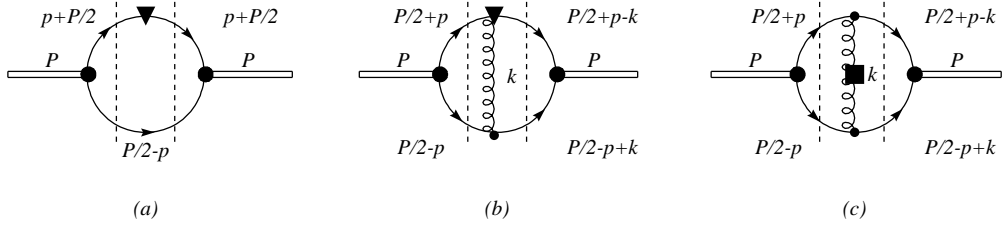}}

\hspace*{1cm}
\begin{minipage}{15cm}
\caption{\em The moments of the deuteron SF up to the second order in 
meson-nucleon coupling: (a) relativistic impulse
approximation with a struck nucleon projected to its mass-shell, (b)
and (c) meson exchange currents. Closed triangles 
and square denote the moments of the free nucleon and meson SF's, 
respectively. The vertical dashed lines separate operators and the
BS vertex function of the deuteron in corresponding matrix elements.}
\end{minipage}\label{diag}
\end{figure}

The first two moments of $\mu_n^{N/D}$ have a physical interpretation. For 
$n=1$ we find that $\mu_1^{N/D}({\rm int})=0$ and as a result the baryon
number associated with the nucleons fields is preserved. The
normalisation condition~(\ref{normcon}) implies that  
$\mu^{N/D}_1=\mu_1({\rm RIA})^{N/D}=1$. For  $n=2$ Eq.~(\ref{mu-int})
contributes to the energy-momentum sum rule. The fraction of the total
momentum of the deuteron carried by nucleons is  
\begin{eqnarray}
\frac{2m}{M}\langle z \rangle=\mu_2^{N/D}({\rm RIA})
+\mu_2^{N/D}({\rm int}),\label{y}
\end{eqnarray}
where at the deuteron rest frame, $P_+=M$, we have
\begin{eqnarray}
&&\mu_2^{N/D}({\rm RIA})=-\frac{\imath}{M^2}\int d^4p\, \tilde p_+ 
\bar\chi(p;P)\gamma_+^{(1)}S^{-1}(p_2)
\chi(p;P),\label{SR-IA}\\
&&\mu_2^{N/D}({\rm int})=-\frac{1}{M^2}\int \frac{d^4pd^4k}{(2\pi)^4}\,
\bar\chi(p;P)\frac{
[\gamma_+\gamma^0]^{(1)}{\cal V}_{{\rm OBEP}}+
{\cal V}_{{\rm OBEP}}[\gamma^0\gamma_+]^{(1)}}{2}
\chi(p+k;P).\label{SR-int}
\end{eqnarray}               

In order to estimate contribution given by Eq.~(\ref{SR-int}) we can 
manipulate with the BSE~(\ref{BS-eq}) and find
\begin{eqnarray}
\mu_2^{N/D}({\rm int})=\frac{\imath}{M^2}\int d^4p\,
\bar\chi(p;P)\frac{[\gamma_+\gamma^0 S^{-1}(p_1)
+
S^{-1}(p_1)
\gamma^0\gamma_+]^{(1)}}{2}
S^{-1}(p_2)
\chi(p;P).\label{alter}
\end{eqnarray}

The nucleon contribution to the deuteron momentum can be written
as $\langle z\rangle=1-\Delta$, where $\Delta$ is the missing fraction
of
the deuteron momentum carried by exchange mesons. Notice that Eq.
(\ref{alter}) suggests that the magnitude of the binding in the 
deuteron does not explicitly depend on the interaction kernel.

\subsection{Meson exchange currents}
\indent

The meson part of the nuclear SF is concentrated at small values of
$x$. We expect that the bulk of the relevant meson effects in deep 
inelastic scattering on the deuteron to be associated with pion
exchange. Therefore, we consider the role of pions only in this section.
Although other kinds of meson exchanges which appear in the OBE kernel
of Eq.~(\ref{BS-eq}) are taken into account in this approach, it is 
impossible to estimate their contribution to deuteron SF.  There is
no data on the free SF's for these mesons. We can also expect that
contribution due to the scalar and vector 
exchange in the deuteron cancels, if both kinds are considered in the 
manner given by the operators in Eqs.~(\ref{sca-S}) and~(\ref{vec-V3}).   

In the impulse approximation, the contribution of the mesons according 
to Fig.~1(c) is summed up to (here we write down expression for  
$\pi$-mesons only):
\begin{eqnarray}
\mu_n^{\pi/D}=\int d^4k\, n_\pi(k;P)\frac{k_+}{\Omega_k}
(\frac{k_+}{P_+})^{n-1},\label{mu-meson}
\end{eqnarray}
where $\Omega^2_k=k^2-\mu_\pi^2$ and function $n_\pi(k)$ can be 
interpreted as the excess pion 4-momentum distribution associated 
with the one-pion exchange in the deuteron  
\begin{eqnarray}
n_\pi(k;P)=\frac{1}{P_+}\int \frac{d^4p}{(2\pi)^4}\,
\bar\chi(p;P)\frac{{\cal V}_{\pi}(k)}{\Omega_k}\chi(p+k;P),\label{pi-dist}
\end{eqnarray}
where the one-pion exchange kernel ${\cal V}_\pi$ appearing in
Eq. (\ref{pi-dist}) is used in solving the BSE~(\ref{BS-eq}). 

The first two moments of $\mu^{\pi/D}_n$ are interpreted as 
 the mean number of the excess 
pions in deep inelastic scattering, 
$\mu^{\pi/D}_1=\langle N_\pi\rangle$, and the missing 
fraction of the deuteron momentum carried by the excess pions, 
$\mu^{\pi/D}_2=\langle y\rangle_\pi$, respectively. Since 
the twist-two operators from the set (b) for $n=2$  are related to the
energy-momentum tensor, the deuteron SF obeys the momentum 
sum rule~\cite{umnikov,birbrair}:
\begin{eqnarray}
\int\limits_0^1F_2^D(x)dx=\frac{1}{P_0}\langle P|
\Theta^{00}+\Theta^{33}|P\rangle=1,\label{smrule}
\end{eqnarray}
where the components of the energy-momentum tensor $\Theta^{\mu\nu}$
 correspond to
those in Eq.~(\ref{tensor}). Eq.~(\ref{smrule}) ensures that the MEC 
carry away the missing part of the deuteron momentum.

\subsection{Non-relativistic limit}
\indent

In order to find the NR of the theory, though it
may seem to be dubious, we need to take into account only the 
positive-energy states
in the BS amplitude. The negative-energy components do not survive
in the limit $|{\bf p}|/m\ll 1$. 

We write down the BS amplitudes in the rest frame of the deuteron
as decomposition in terms of the product of Dirac spinor of nucleons:
\begin{eqnarray}
&&\chi(p;P_{(0)})=u({\bf p})u(-{\bf p})
\bigl[{\cal Y}_{1\cal M}^{01}(\hat {\bf p})u(p_0,|{\bf p}|)+
{\cal Y}_{1\cal M}^{21}(\hat {\bf p})w(p_0,|{\bf p}|) \bigr],\label{wf}
\end{eqnarray}
where $u({\bf p})$ is the positive-energy Dirac spinor. 
 We rewrite Eq.~(\ref{wf}) equivalently as   
\begin{eqnarray}
&&\chi(p;P_{(0)})=u({\bf p})u(-{\bf p})\Psi(p_0,{\bf p}),\\
&&\bar\chi(p;P_{(0)})=-\bar u({\bf p})\bar u(-{\bf p})\Psi^*(-p_0,{\bf
p}),\nonumber
\end{eqnarray}
where the radial wave functions $u(p_0,|{\bf p}|)$ and $w(p_0,|{\bf
p}|)$ can be expressed in terms of the BS vertex functions 
$\Gamma_L$ with $L=0,2$ as follows ($E_{\bf p}=\sqrt{{\bf p}^2+m^2}$)
\begin{eqnarray}
u(p_0,|{\bf p}|)=\frac{1}{\sqrt{N}}
\frac{\Gamma_0(0,|{\bf p}|)\xi_0(p_0,|{\bf p}|)}
{\bigl[(\frac{M}{2}-E_{\bf p})^2-p_0^2\bigr]},\quad 
w(p_0,|{\bf p}|)=\frac{1}{\sqrt{N}}
\frac{\Gamma_2(0,|{\bf p}|)\xi_2(p_0,|{\bf p}|)}
{\bigl[(\frac{M}{2}-E_{\bf p})^2-p_0^2\bigr]}\label{uw}
\end{eqnarray}
with a smooth function $\xi_L(p_0,|{\bf p}|)$ reflecting the 
dependence on the relative energy $p_0$ of $\Gamma_L$.  Let us 
consider Eq.~(\ref{mu-IA}) for the positive-energy components 
only. We find that 
\begin{eqnarray}
\mu_n^{N/D}({\rm RIA})=\frac{\imath}{M}\int d^4p\,
|\Psi(p)|^2\bigl(1+\frac{p_z}{E_{\bf p}}\bigr)
(E_{\bf p}-p_2^0)(\frac{\tilde p_+}{M})^{n-1},\label{mu-IA1}
\end{eqnarray}
where $\tilde p_+^\mu=(E_{\bf p},{\bf p})$ and ${p_2}^\mu=(M_d/2-p_0,-{\bf
p})$. Our next step is to integrate over the relative energy picking up 
the pole in the 4-quarter:
\begin{eqnarray}
\int\limits_{-\infty}^{+\infty} \frac{dp_0}{\imath 2\pi}
\frac{\xi_L(p_0,|{\bf p}|)\xi_{L^\prime}(p_0,|{\bf p}|)}
{(\frac{M}2-E_{\bf p}-p_0+\imath\epsilon
)(\frac{M}2-E_{\bf p}+p_0+\imath\epsilon)^2}=
\frac{\xi_L(p)\xi_{L^\prime}(p)}
{(\frac{M}2-E_{\bf p}-p_0)^2}\Biggl|_{p_0=E_{\bf
p}-\frac{M}2},
\label{pole}
\end{eqnarray}
Keeping in mind Eq.~(\ref{pole}) we reduce Eq.~(\ref{mu-IA1}) to the
form~\cite{new}
\begin{eqnarray}
\mu_n^{N/D}({\rm IA})=
\int d{\bf p}\, n({\bf p})\bigl(1+\frac{p_z}{E_{\bf p}}\bigr)
(\frac{E_{\bf p}+p_z}{M})^{n-1}\label{mu-IAnr}
\end{eqnarray}
with the nucleon density $n({\bf p})=|\Psi|^2$ depending 
on radial wave functions of the $L=0,2$ angular momentum states
($\xi\approx 1$)
\begin{eqnarray}
\sqrt{\frac{2\pi}{N}}\frac{
\Gamma_L(0,|{\bf p}|)}
{\sqrt{M}(2E_{\bf p}-M)},
\end{eqnarray} 
which in the NR limit are usually identified with the 
conventional wave functions in $^3S_1$- and $^3D_1$-states. 
Eq.~(\ref{mu-IAnr}) is the impulse approximation   with a hit
nucleon on its mass-shell. 

The other important quantity is $\langle z\rangle$, because its magnitude
is well-known from NR studies. First of all we integrate over $p_0$ in
Eq.~(\ref{alter}) in the same manner as done above. We get 
\begin{eqnarray}
\mu_2^{N/D}({\rm int})=\frac{1}{M}\int d{\bf p}\,
n({\bf p})(M-2E_{\bf p})=
\frac{2}{M}(M-2\langle E_{\bf p}\rangle).
\end{eqnarray} 
Using Eq.~(\ref{mu-IAnr}) for $n=2$ we obtain
\begin{eqnarray}
\mu_2^{N/D}({\rm RIA})=\frac{2}{M}(\langle E_{\bf p}\rangle+\frac{\langle
p_z^2\rangle}{E_{\bf p}}).
\end{eqnarray}

It is clear that the proton and neutron in the deuteron carry the
fraction of the total momentum of the deuteron given by~\cite{physrev}
\begin{eqnarray}
\langle z\rangle=\frac{1}{m}(M-\langle E_{\bf p}\rangle+\frac{\langle
p_z^2\rangle}{E_{\bf p}})\approx 
\underbrace{1+\frac{5\langle T\rangle}{6m}}_{{\rm IA}}
+\underbrace{\frac{\langle V\rangle}{m}}_{{\rm int}},
\end{eqnarray}
where $\langle T\rangle=\langle {\bf p}^2\rangle/m$ is the kinetic 
energy, $\langle V\rangle=-\langle T\rangle+\varepsilon_d$ is the 
potential energy of two nucleons in the deuteron and $\varepsilon_d=
M-2m$.

\section{Results and discussions}
\indent 

Owing to Eq.~(\ref{mom-conv}) the SF $F_2^D(x,Q^2)$ 
is written as a sum of convolutions of the SF of the unbound nucleon and 
pion and the momentum distribution functions of the nucleons and 
pions inside the deuteron (we assume that the SF $F_2^{N,\pi}$ do 
not carry any dependence on the relative 4-momentum). The contribution
of the interaction corrections to the RIA given by Eq.~(\ref{mu-int}) 
can be approximated in a very plain form yielding  
\begin{eqnarray}
F_2^{D}(x,Q^2)=F_2^{N/D}(x,Q^2)-
\frac{\langle {\cal V}\rangle}{m}
\frac{\partial}{\partial
{\rm ln} x}F_2^N(x,Q^2)+
\delta F^{\pi/D}(x,Q^2),\label{Fd}
\end{eqnarray}  
where $\langle {\cal V}\rangle=m\mu_2^{N/D}({\rm int})$ and 
\begin{eqnarray}
F_2^{N/D}(x,Q^2)=\int\limits_{z\ge x}f^{N/D}(z)F_2^N(x/z,Q^2)dz,\label{Fdn}\\
\delta F^{\pi/D}(x,Q^2)
=\int\limits_{y\ge x}f^{\pi/D}(y)F_2^\pi(x/y,Q^2)dy.\label{Fdpi}
\end{eqnarray}

The distribution functions $f^{N/D}(z)$ and $f^{\pi/D}(y)$
describe the momentum distributions of the on-mass-shell nucleons 
and off-mass-shell pions carrying the fractional longitudinal momenta
of the deuteron $y$ and $z$, respectively. We have at the rest frame: 
\begin{eqnarray}
&& f^{N/D}(z)=\frac{1}{M}\sum_\alpha
\int d^4p\,
|\phi_\alpha (p;P_{(0)})|^2\left(1+\frac{p_z}{E_{\bf p}}\right)
\delta\left(z-\frac{{\tilde p}_+^\alpha}{M}\right),\label{fnd}\\
&& f^{\pi/D}(y)=
\int d^4k\, n_\pi(k;P_{(0)})\frac{k_+}{\Omega_k}
\delta\left(y-\frac{k_+}{M}\right),\label{fpid}
\end{eqnarray}
where ${\tilde p}^\alpha=(\rho E_{\bf p},{\bf p})$ is the momentum of the 
struck nucleon in the partial state $\alpha={}^{2S+1}L_{J=1}^{\rho}$; 
$\phi_\alpha$ denotes a partial amplitude. 

Both functions $f^{N/D}(z)$ and $f^{\pi/D}(y)$ specify the 
number densities of the nucleons and excess pions and 
satisfy the normalisation integrals 
\begin{eqnarray}
\int f^{N/D}(z)dz=2,\quad \langle N_\pi\rangle=\int f^{\pi/D}(y)dy.
\end{eqnarray}
The SF $F_2$ probes the momentum distribution of the constituents and
thus conservation of the energy-momentum requires that 
\begin{eqnarray}
\int f^{N/D}(z)zdz+\int f^{\pi/D}(y)ydy+\int f^{B/D}(y)ydy=1,
\label{f-sr}
\end{eqnarray}
where $f^{B/D}(y)$ the momentum distributions of some other mesons in
${\cal V}_{\rm OBEP}$.

For the sake of a simple analytical analysis of the nucleon density in
Eq.~(\ref{fnd}) and leaving no doubts of accuracy of theoretical 
predictions, we employ the BS vertex function which is the solution of 
the homogeneous BSE  for two spin-$\frac12$ particles interacting  
through a covariant, separable potential~\cite{rupp}. The relevance of
such an approximation to the expressions with the original BS vertex
function of the OBE potential could be justified. Indeed the final formulae 
for $\mu_2^{N/D}({\rm int})$, $\langle z\rangle$ and $\langle
N_B\rangle$ (v.~i.) do not contain the OBE potential. On the other 
hand discarding of the negative-energy components in the vertex function 
is also justified by the conclusion in Ref.~\cite{rupp}. 

The magnitude of the binding in the deuteron can be calculated in a 
straightforward
way by making use of Eq.~(\ref{alter}). Computation of 
$\langle z\rangle$ in the energy-momentum sum rule gives: the 
on-mass-shell nucleons carry $\langle z\rangle_{\rm RIA}=1.0153$ which
breaks down the sum rule~(\ref{f-sr}). The interaction term~(\ref{mu-int})
takes away a fraction of the longitudinal momentum, 
$\mu_2({\rm int})=-0.0188$, and reduces $\langle z\rangle_{\rm RIA}$ 
yielding the final value, $\langle z\rangle=0.9965$. 

Unfortunately the total contribution of the mesons cannot be
calculated explicitly without running into formidable computation. Yet
 we find by means of Eq.~(\ref{f-sr}) that the fraction of the 
momentum carried by the exchange 
mesons is $\Delta=0.0035$. The first moment $\mu_1^{B/D}$
is the mean number of the exchange mesons associated with the 
reaction. The exact 
expression being similar to Eq.~(\ref{mu-meson}) can be approximated by the 
formula
\begin{eqnarray}
\langle N_B\rangle\approx
\frac{\imath}{2P_+}\int d^4p\,\bar\chi(p;P)
\frac{\partial}{\partial p_-}S^{-1}(p_1,p_2)\chi(p;P),
\end{eqnarray}
where $p_-=p_0-p_z$. It is expected that $\langle N_B\rangle$ is totally
accounted for by the pions, since the part due to heavier meson
exchanges is suppressed in average by their larger masses in the propagator
$\Omega_k^2$ and their effective couplings. If we pretend not to take 
interest
in a detail structure of the $f^{\pi/D}(y)$ that emerges from
Eq.~(\ref{fpid}),  we can tentatively 
estimate the momentum distribution of the pions for not sufficiently
large values of $z$ by exploiting results of Ref.~\cite{berger}. The
quantitative analysis shows that results are not sensitive to the
precise functional shape of $f^{\pi/D}(y)$ but controlled mainly by 
$\langle N_\pi\rangle$ and $\langle y\rangle$. Thus  $f^{\pi/D}(y)$
could be parametrised imposing the balance between $\langle
N_\pi\rangle$ and $\langle y\rangle_\pi$ in the simplest manner. 
  
\begin{figure}[t]
\input epsf
\epsfxsize=7.5cm
\epsfysize=9cm
\hspace*{0.5cm}\epsfbox{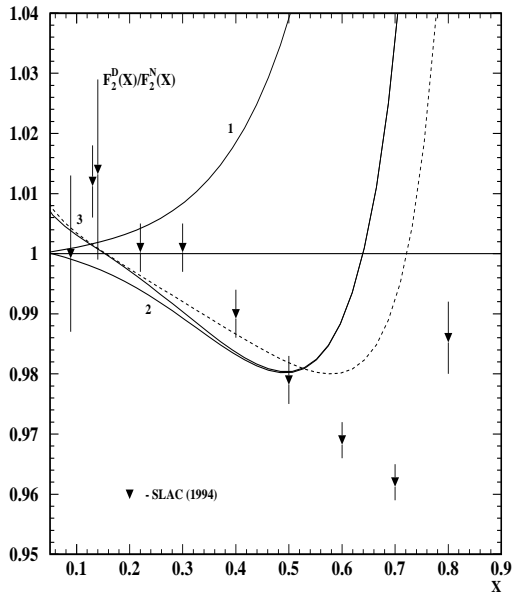}

\vspace*{-8cm}

\hspace*{8.0cm}
\begin{minipage}{8cm}
\caption{\em The ratio of the deuteron and isoscalar nucleon 
structure functions. Solid curves: curve 1 denotes the relativistic 
impulse approximation (RIA) with on-mass-shell kinematics; curve 2 
denotes the RIA with taking into account the nuclear binding; curve 3
denotes the sum contribution of the RIA, the binding and
meson exchange currents. The
dashed curve corresponds to the solid curve 3 with a different 
parametrisation of $F_2^N(x,Q^2)$. Experimental data are taken from 
Ref.~\protect{\cite{gomez}}. The error bars are combined statistical and 
systematic errors.}

\vspace*{1cm}

\end{minipage}
\end{figure}

\subsection{Comparison with data}
\indent 

Now we are in  a position to perform numerical analysis.
Fig.~2 shows the EMC effect in the deuteron at $Q^2=25$~GeV$^2$ given
by the ratio $R(x,Q^2)=F_2^D(x,Q^2)/F_2^N(x,Q^2)$ in
the present approach. 
All computations  of the nucleon momentum distribution $f^{N/D}(z)$ and
$\mu_2(int)$ are carried out using the BS vertex function in the
framework with multi-rank 
separable interaction~\cite{rupp}. The free nucleon 
$F_2^N=(F_2^p+F_2^n)/2$ 
and pion $F_2^\pi(x,Q\approx 25~{\rm GeV}^2)$ SF's  obtained from the 
combined proton and deuteron data of the BCDMS collaboration~\cite{bcdms} 
and data on massive lepton pair production of the CERN NA10 
collaboration~\cite{na10}, respectively, are taken from 
Refs.~\cite{physrev} and~\cite{berger} .

At {\em small} $x$, $x\le 0.2$, the MEC increase the deuteron SF 
relative to the nucleon one. At exactly $x=0$ the ratio has maximum, 
$R(0)=1+\langle N_\pi\rangle F_2^\pi(0)/F_2^N(0)$, where
$\langle N_\pi\rangle\approx0.015$, and falls off roughly linearly as $x$
grows. The value of $F_2^\pi(0)/F_2^N(0)$ is known from the 
parametrisation of the SF's. As can be seen, the nucleon contribution 
$F^{N/D}(x,Q^2)$ to $R(x,Q^2)$ is not important in this range of $x$. 

At {\em intermediate} $x$, $0.2<x<0.6$, the MEC are negligible as long as  
$f^{\pi/D}(y)$ vanishes at $y\to 1$, but the correction due to strong 
interaction between nucleons, v.~s.  
Eq.~(\ref{Fd}),  becomes responsible for the dip. This depletion is
totally controlled by the value of $\mu_2({\rm int})$.  In a more realistic
case we still may have contribution from deep inelastic scattering on
pions in this region, because the argument $x/y$ of $F_2^\pi$ in the 
convolution formula, see Eq.~(\ref{Fdpi}), will run over broader range
in $f^{\pi/D}(y)$.

The rise of the ratio $R(x)$ at {\em large} $x$, $x>0.6$, is associated with
the relativistic Fermi motion. The tendency is that the harder ``tail'' of 
$F_2^D(x)$ in the vicinity of the boundary for the single-nucleon
kinematics, the harder the momentum distribution $f^{N/D}(z)$ in 
Eq.~(\ref{Fdn}). 

Unfortunately pertinent data on the ratio $R(x,Q^2)$ are not still available. 
However in Ref.~\cite{frankfurt} it is 
suggested that the $A$-dependence of the 
$F_2^A(x)$ should be determined by the local properties of nuclear
matter, i. e. the average nuclear density $\rho(A)$. 
The naive extrapolation
to the deuteron case leads to $R(x,Q^2)\approx 1+(F_2^A/F_2^D-1)/(
\rho(A)/\rho(D)-1)$, although this (may) overestimate the effect due to 
the isoscalarity of the deuteron. The model-dependent value of the ratio
$F_2^D/F_2^N$ estimated from averaging over measured data on various 
nuclei $F_2^A/F_2^D$ in SLAC-E139 experiment is presented on Fig.~2. 
Within this hypothesis the deuteron has a significant EMC effect which
contradicts to our theoretical predictions together with the
parametrisation of the neutron SF~\cite{physrev}: a deviation of
$R(x,Q^2)$ from unity at intermediate $x$ is extremely large, 
about 4~\% versus  the theoretically predicted 2~\%, and centers 
of data points continue to decrease up to $x=0.7$ which is the 
feature of data for heavy nuclei. 

Our computation analysis shows that  $R(x,Q^2)$ does not exhibit
substantial dependence on $Q^2$ in the broad range and on a 
specific parametrisation of the neutron SF (both effects are of the
second order relative to the binding).  The ratio is more sensitive to
the behaviour of the free nucleon SF as $x\to 1$, since this region is
enhanced by the nucleon momentum distribution. 

\section{Concluding remarks}
\indent 

In this paper the deuteron SF  $F_2(x,Q^2)$ is
examined in the 
theoretical approach based on a picture in which nucleons and
mesons are the relevant degrees of freedom. The
deuteron is treated as two spin-$\frac12$ particle bound state
through a consideration of the Bethe-Salpeter equation in the ladder 
approximation. In the model the
deuteron SF given by Eq.~(\ref{Fd}) is obtained as a sum of
three terms. The first term known as the relativistic impulse
approximation is 
the convolution of the the free nucleon SF with the fractional
momentum distribution function of the nucleons 
projected to their mass-shell. The second term in
Eq.~(\ref{Fd}) accounts for the binding correction to the
 relativistic 
impulse approximation. Similarly to the first term the last piece is
the convolution of the free pion SF with
the excess pion fractional momentum distribution in the deuteron. The 
momentum 
distributions of the deuteron constituents are derived from the matrix
elements
of certain twist-two operators in the operator product expansion
method and expressed in terms of the BS vertex function.
 
The principal result of the paper is that the binding in the
deuteron in the approach incorporating the relativistic dynamics in
terms of the conventional 
 degrees of freedom could be revealed in an unambiguous way. 
The relativistic impulse approximation 
is obtained in the form that avoids the explicit dependence 
of the momentum distribution on the relative energy of
nucleons and  thus we are able to present accurate estimates in a
manner similar to the non-relativistic treatment of Ref.~\cite{new}. 

The magnitude of the nuclear binding and Fermi motion correction 
together with a neutron SF $F_2^n$ shape $x$-dependence of the 
EMC ratio for the deuteron. High precision measurements of the 
nucleon SF at large $x$ combined with a reliable theoretical treatment
 of the
deuteron may help to resolve the question of the 
extraction of the neutron SF from experimental data on a deuterium  
target and a heavy nuclei target~\cite{luiti}.

The open questions the model suggests are that an additional work to be 
done in order to establish connection between the off-shell and
on-shell SF $F_2^N$ in this framework; next to what extent the
negative-energy states in 
the BS vertex function impact on the momentum distributions of the 
nucleons; and finally the exact functional shape of the momentum 
distribution of the excess pions associated with the one pion exchange 
in the deuteron is not of the least interest.    
 
To sum up, we can conclude that the point of view developed in this 
paper might help to discriminate the contribution of the pure binding 
effects in the nuclear SF and regard  corrections due to the 
relativistic Fermi motion in a transparent way.  

\vskip 5mm 

\centerline{\large \bf Acknowledgements}
\vskip 2mm
\indent 

It is a pleasure for us to express thanks to F.C. Khanna and A.V. Molochkov
for stimulating discussions on the subject of the paper. We are gratefully
acknowledge helpful conversations with I.N. Mishustin.  
We also thank S.M. Dorkin and S.G. Bondarenko for consulting on 
numerics. 

This work was performed under the financial support from  the Danish 
Ministry of Education and supported in part by the NSERC of Canada. 
One of us (K.Yu.K.) is thankful to the Niels Bohr Institute  for the 
warm hospitality and computer facilities.


 
\end{document}